\documentclass[twocolumn,showpacs,preprintnumbers,amsmath,amssymb,prb,superscriptaddress]{revtex4-2}
\usepackage[dvipdfmx]{graphicx}
\usepackage{dcolumn}
\usepackage{bm}
\usepackage{color} 
\usepackage{ulem} 

\usepackage{hyperref}
\hypersetup{
  colorlinks   = true,
  urlcolor     = blue, 
  linkcolor    = blue, 
  citecolor   = blue
}
\usepackage[all]{hypcap}
\usepackage{xspace}

\newcommand{\TC}{$T_{\rm C}$\xspace}

\newcommand{\EF}{$E_{\rm F}$\xspace}

\newcommand{\CeAlSi}{CeAlSi\xspace}
\newcommand{\LaAlSi}{LaAlSi\xspace}
\newcommand{\DFT}{DFT\xspace}
\newcommand{\DFTU}{DFT+$U$\xspace}
\newcommand{\DFTDMFT}{DFT+DMFT\xspace}

\newcommand{\OC}{$\sigma_1(\omega)$\xspace}

\newcommand{\R}{$R(\omega)$\xspace}
\newcommand{\hw}{$\hbar\omega$\xspace}

\begin{document}

\title{
Optical investigation of the electronic structure of a ferromagnetic Weyl semimetal \CeAlSi
}
\author{Shin-ichi~Kimura}
\email{sk@kimura-lab.com}
\affiliation{Graduate School of Frontier Biosciences, The University of Osaka, Suita, Osaka 565-0871, Japan}
\affiliation{Department of Physics, Graduate School of Science, The University of Osaka, Toyonaka, Osaka 560-0043, Japan}
\affiliation{Institute for Molecular Science, 
Okazaki, Aichi 444-8585, Japan}
\author{Yue~Pan}
\affiliation{Department of Physics, Graduate School of Science, The University of Osaka, Toyonaka, Osaka 560-0043, Japan}
\author{Hiroshi~Watanabe}
 \altaffiliation[Present affiliation: ]{Institute for Chemical Research, Kyoto University, Uji, Kyoto 611-0011, Japan}
\affiliation{Graduate School of Frontier Biosciences, The University of Osaka, Suita, Osaka 565-0871, Japan}
\affiliation{Department of Physics, Graduate School of Science, The University of Osaka, Toyonaka, Osaka 560-0043, Japan}
\author{Akimitsu~Kirikoshi}
\affiliation{Research Institute for Interdisciplinary Science, Okayama University, Okayama 700-8530, Japan}
\author{Junya~Otsuki}
\affiliation{Research Institute for Interdisciplinary Science, Okayama University, Okayama 700-8530, Japan}
\author{Hiroshi~Tanida}
\affiliation{Liberal Arts and Sciences, Toyama Prefectural University, Imizu, Toyama 939-0398, Japan}
\date{\today}
\begin{abstract}
To investigate electronic states during the ferromagnetic transition in a magnetic Weyl semimetal \CeAlSi, we measured temperature-dependent optical conductivity [$\sigma_1(\omega)$] spectra and compared them with \DFTDMFT band calculations.
The \OC spectrum did not change significantly across the ferromagnetic ordering temperature (\TC), suggesting that the Ce~$4f$ states are almost localized.
\DFTDMFT calculations with almost localized Ce~$4f$ states successfully reproduced the spectral shape and the unchanged \OC spectra across \TC.
The dynamic effective mass evaluated from the extended Drude model is very small, which \DFTDMFT calculations also reproduce, but the scattering probability at even lower temperatures suggests ferromagnetic fluctuations.
These results suggest that the interaction intensity between the Weyl fermions and Ce~$4f$ states is very weak, as reproduced by \DFTDMFT calculations.
\end{abstract}

%
\maketitle
%
%
Our understanding of electronic phases based on topology has advanced rapidly in recent years, giving rise to a diverse range of physical properties, not only surface edge states in topological insulators but also the emergence of Dirac semimetals in bulk electronic states, as well as Weyl semimetals appearing in broken spatial or time-reversal symmetries~\cite{Armitage2018-gn}.
In particular, in Weyl semimetals, the spin-nondegenerate conduction and valence bands make linear contact at discrete points, namely Weyl points, and excitations in their vicinity are described by relativistic Weyl fermions.
The Weyl point acts as a mono-magnetic pole in momentum space and gives rise to various anomalous transport phenomena, such as the anomalous Hall effect and negative magnetoresistance~\cite{Cho2023-fs}.
Furthermore, there are typical surface states, namely ``Fermi arcs'', which show opened Fermi surfaces connecting Weyl points in the bulk, resulting in characteristic topological electronic states in both the bulk and at the surface~\cite{Lv2021-vy}.

Weyl semimetals are broadly classified into two categories based on the type of symmetry breaking.
One category consists of non-centrosymmetric Weyl semimetals with broken spatial inversion symmetry; the TaAs family (TaAs, NbAs, TaP, NbP) has been extensively studied~\cite{Yan2017-ox}.
The other category consists of magnetic Weyl semimetals with broken time-reversal symmetry; typical examples include Co$_3$Sn$_2$S$_2$~\cite{Wang2018-qe} and Mn$_3$Sn~\cite{Kimata2019-bl}.
\CeAlSi, the subject of this study, is a non-centrosymmetric tetragonal structure (LaPtSi-type structure) belonging to the space group $I4_{1}md$ (No.~109)~\cite{Pottgen2016-sn}.
Furthermore, it is an extremely rare candidate for a Weyl semimetal that also possesses a non-collinear ferromagnetic (FM) structure with a ferromagnetic ordering temperature of \TC$\sim9~{\rm K}$ due to localized spins of rare-earth elements~\cite{Sun2021-vw}.
The localized magnetism originating from the $4f$ electrons of Ce atoms is thought to strongly influence the electronic band structure and transport properties through interactions with conduction electrons; furthermore, an anomalous Hall effect has been observed in this material~\cite{Alam2023-na,Cheng2024-hw,Zang2025-md}.
A major characteristic of this material is the coexistence of breaks in spatial inversion symmetry and in time-reversal symmetry due to magnetic order.
It has been suggested that such combined symmetry breaking could give rise to a complex and distinctive topological electronic structure that is not observed in other Weyl semimetals.

In recent years, multifaced research on \CeAlSi has been conducted, including angle-resolved photoelectron spectroscopy (ARPES) measurements~\cite{Sakhya2023-en,Cheng2024-hw,Morita2025-ek}, magnetic transport measurements~\cite{Zang2025-md}, and quantum oscillations (QO)~\cite{Meena2025-pj}.
In particular, low-energy ARPES experiments with vacuum-ultraviolet photons clearly show the Fermi arc~\cite{Sakhya2023-en} and its strong dependence on magnetic ordering~\cite {Cheng2024-hw}.
On the other hand, high-energy ARPES experiments with soft-X-ray photons revealed the bulk band structure, explained with \DFT calculations of \LaAlSi and \DFTU calculations of \CeAlSi, where the Ce~$4f$ states are treated as fully localized~\cite{Morita2025-ek}.
However, the effect of FM ordering on the bulk electronic structure at zero magnetic field has not yet been revealed because of the limited measurement temperature range, although a QO measurement suggests that FM ordering modifies the Fermi-surface topology at high magnetic field~\cite{Piva2023-ij}.

Regarding its optical response, there have been reports on the $R$-dependence of $R$AlSi ($R$: rare earth), including \CeAlSi~\cite{Kunze2024-qf}, in the paramagnetic (PM) phase at room temperature.
In a related material, {PrAlSi}, changes in the optical spectrum associated with the FM transition have been observed~\cite{Gao2025-pq}.
However, in \CeAlSi, changes in the Weyl electron states accompanying the magnetic transition, and comparisons with theoretical calculations, remain to be clarified.

In this study, we compared the \OC spectra of \CeAlSi with band calculations based on \DFT and \DFTDMFT to determine the effect of the Ce~$4f$ states on the low-energy electronic states of \CeAlSi.
Furthermore, we investigated changes in the \OC spectra at \TC and, by comparing with band calculations, demonstrated the nature of the electronic states formed.

%
%
%
\begin{figure}[t]
\begin{center}
\includegraphics[width=0.42\textwidth]{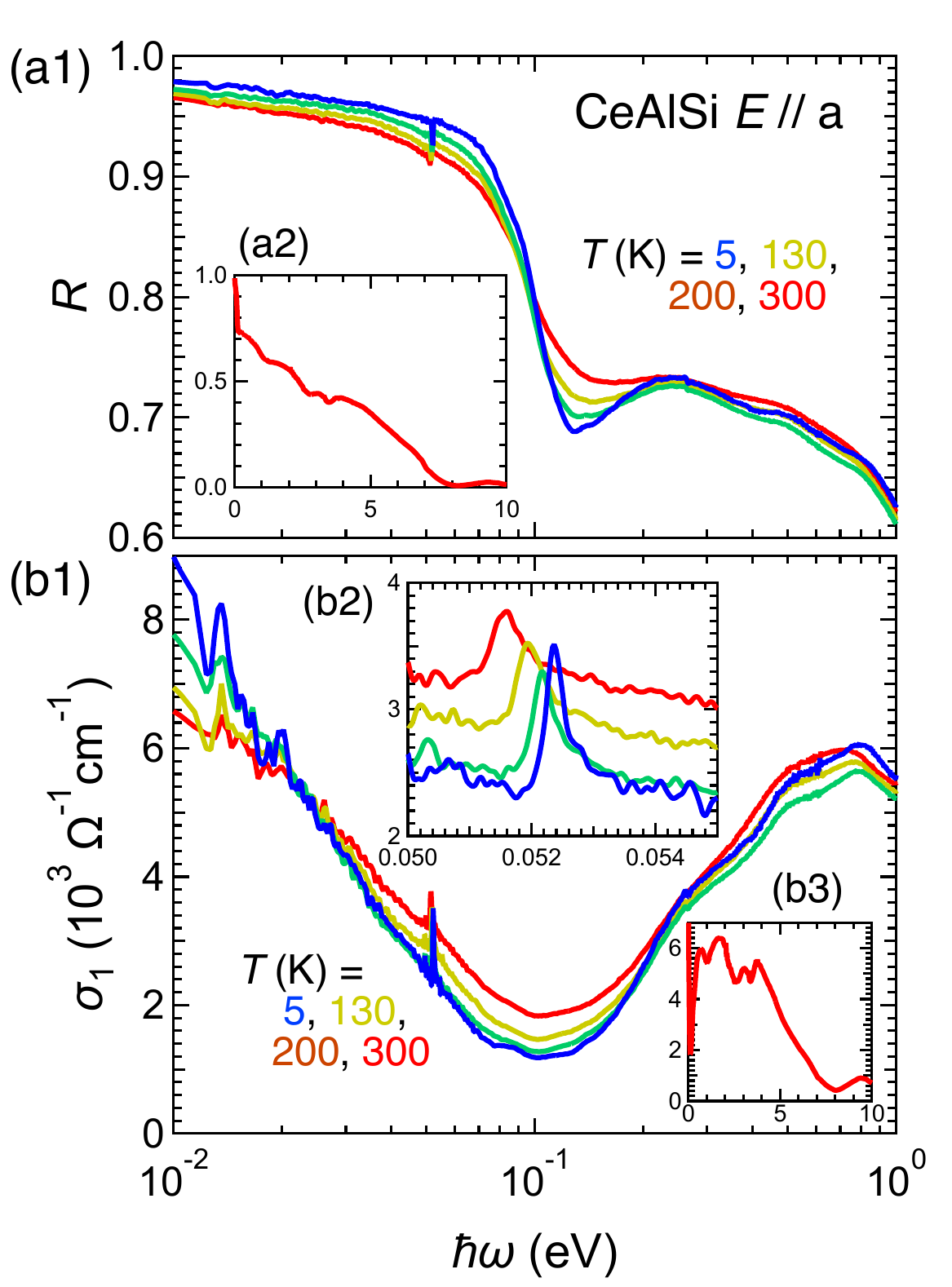}
\end{center}
\caption{
(a1) Temperature-dependent reflectivity [$R(\omega)$] spectra of CeAlSi with the electric vector parallel to the $a$ axis ($E \parallel a$) in the photon energy $\hbar\omega$ range of $0.01-1$~eV.
(a2) Wide-range $R(\omega)$ spectrum up to $10$~eV at 300~K.
(b1) Temperature-dependent optical conductivity [$\sigma_1(\omega)$] spectra of CeAlSi with the same horizontal scale as (a1).
(b2) Enlarged \OC spectra for a TO-phonon peak at $\hbar\omega$ $\sim0.052$~eV.
(b3) Wide-energy-range \OC spectra up to 10~eV at 300~K.
}
\label{fig:reflectivity}
\end{figure}

Single crystals of \CeAlSi and \LaAlSi were grown by the aluminum-self-flux method using an alumina crucible, sealed in a silica tube under an Ar atmosphere.
The temperature-dependent \OC spectra were obtained from the Kramers-Kronig analysis (KKA)~\cite{Kimura2013-rg} of the near-normal-incident polarized optical reflectivity [$R(\omega)$] spectra along the $a$-axis ($E\parallel a$) with a VUV spectrum using a synchrotron beamline~\cite{Fukui2014-wz,Ota2022-ak} and appropriate extrapolations~\cite{Dressel2002-ae} with an experimental direct current conductivity~\cite{Piva2025-oj}.
The detailed explanation is shown in the supplementary materials (SM)~\cite{SM}
Obtained temperature-dependent \R spectra of \CeAlSi are shown in Fig.~\ref{fig:reflectivity}(a1), and the wide-energy-range \R spectrum at 300~K is shown in Fig.~\ref{fig:reflectivity}(a2).

The temperature dependence of the \OC spectrum, obtained by applying  KKA to the \R spectrum, is shown in Fig.~\ref{fig:reflectivity}(b1-b3).
Fig.~\ref{fig:reflectivity}(b3) shows the \OC spectrum corresponding to interband transitions; its low-energy side is shown in Fig.~\ref{fig:reflectivity}(b1), and the drop in \OC of $\sim 0.1~{\rm eV}$ corresponds to the plasma edge caused by carriers.
This energy is about one order of magnitude lower than that of a typical metal, and since the carrier density is proportional to the square of the plasma frequency, the carrier density is on the order of $1~\%$ per unit cell~\cite{Piva2023-ij}.
This supports the conclusion that the carriers originate from the low density of states near Weyl points.
Furthermore, Fig.~\ref{fig:reflectivity}(b2) shows the temperature dependence of optical phonons; the asymmetric shape of the phonon peak, which is characteristic of the Fano effect~\cite{Fano1961-kl}, suggests an interaction with carriers.

%
First-principles \DFT calculations of the band structure have been performed using the {\sc Wien2k} code, including spin-orbit interaction~\cite{Blaha2020-vn}, to explain the experimental \OC spectra.
We used lattice parameters reported in Ref.~\cite{Yang2021-ht} for the calculations.

The \DFTDMFT calculations were performed as follows:
The electronic structure was calculated using the full-potential local-orbital (FPLO) code~\cite{Koepernik1999,Koepernik2023}.
We constructed a tight-binding model consisting of Ce~$4f$ and $5d$, Al~$3s$, $3p$, and $3d$, and Si~$3s$, $3p$, and $3d$ orbitals.
The local Coulomb interactions among the Ce~$4f$ electrons were described by the standard parametrization~\cite{Anisimov1997}, with the Coulomb repulsion $U=6.30~\mathrm{eV}$ and Hund's coupling $J_{\mathrm{H}}=0.79~\mathrm{eV}$.
These parameters, together with the $4f$ energy level $\varepsilon_f=-2.10~\mathrm{eV}$, were chosen to reproduce the $4f^0$ and $4f^2$ excitation energies of elemental Ce~\cite{Herbst1976,Herbst1978,Locht2016} within the Hubbard-I approximation (HIA) that corresponds to the fully localized limit of the $4f$ electrons.
We solved the effective impurity problem in DMFT by the hybridization-expansion continuous-time quantum Monte Carlo (CTQMC) method~\cite{Werner2006,Gull2011}, retaining only density-density interactions.
The self-consistent DMFT calculations were performed with DCore~\cite{Shinaoka2021}. 
The FM state was described by introducing a molecular field consistent with the experimentally observed magnetic structure~\cite{Yang2021-ht}.
The resulting single-particle spectral functions $A(\bm{k},\omega)$ and $A(\omega)$ for the PM and FM phases are shown in Fig.~S2 in SM~\cite{SM}.
The Weyl-point search was performed using WannierTools~\cite{WU2018} after constructing an effective Hamiltonian that includes local correlations within the Hubbard-I approximation~\cite{kirikoshi2026}.



\begin{figure}[t]
\begin{center}
\includegraphics[width=0.40\textwidth]{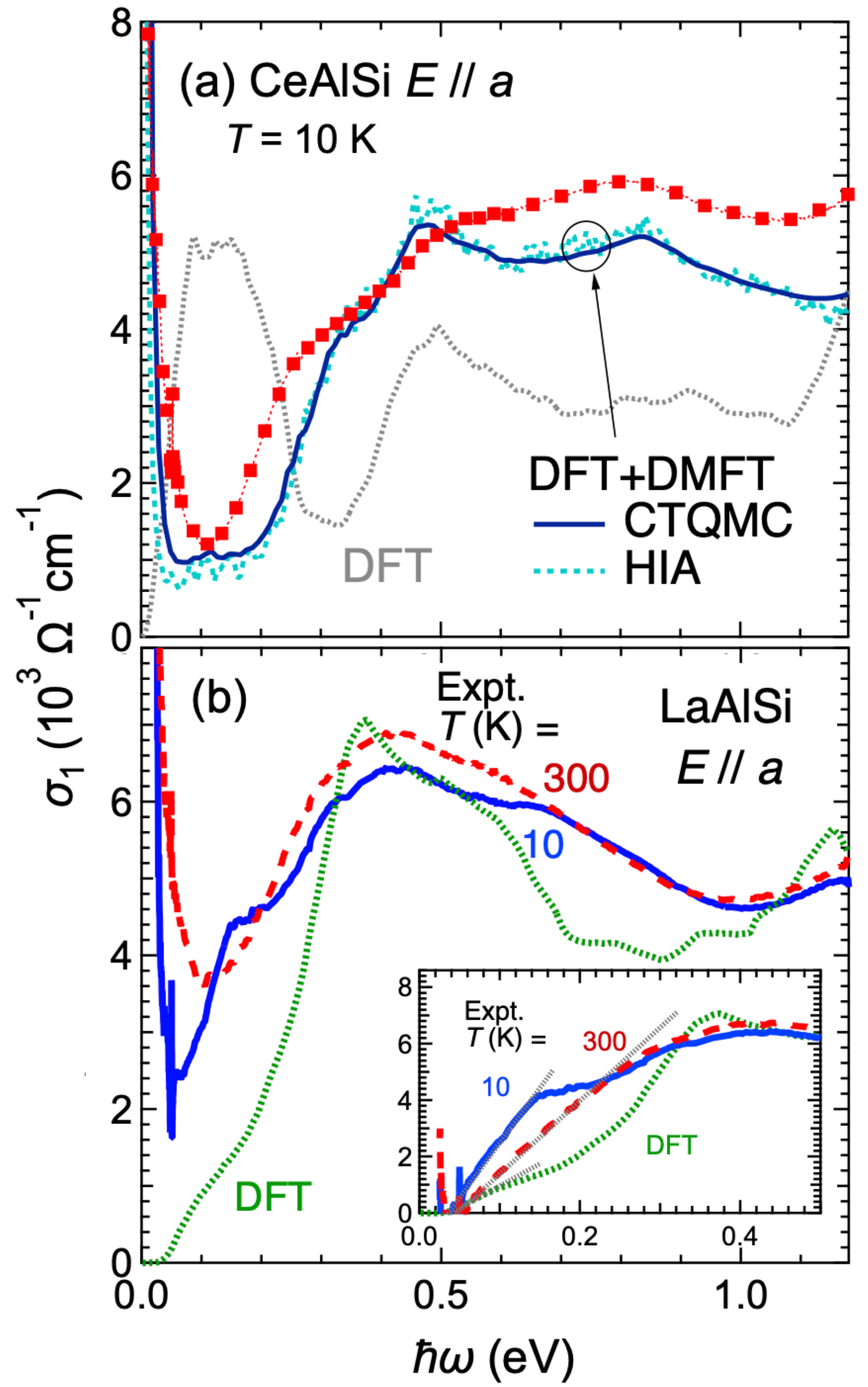}
\end{center}
\caption{
(a) Experimental \OC spectrum of \CeAlSi along the $a$-axis at $T = 10~{\rm K}$ of the PM phase (solid squares) in comparison with calculated spectra with \DFT (dashed line), \DFTDMFT with CTQMC (solid line), and \DFTDMFT with HIA (dotted line).
Note that the \OC spectra with \DFT only indicate the interband components, but the spectra with two \DFTDMFT's includes the Drude component.
(b) Experimental \OC spectra of \LaAlSi along the $a$-axis at $T = 10~{\rm K}$ (solid line) and 300~K  (dashed line) compared with the \DFT calculated \OC spectrum (dotted line).
(Inset) Extended figure near the onset of the interband transition in the \hw region below 0.5~eV.
The experimental spectra are shown after subtracting Drude components.
Gray linear lines are linear components corresponding to the interband transition in Weyl bands.
}
\label{fig:OC}
\end{figure}

First, we examine in detail the structure in the infrared region shown in Fig.~\ref{fig:reflectivity}(b1).
To check whether the experimentally obtained spectral feature in the interband transition part in the \hw range above 0.1~eV can be reproduced with band calculations, we compare the experimental \OC spectra with those obtained from band calculations.
Figure~\ref{fig:OC} shows experimental \OC spectra of \CeAlSi (a) in the PM state at 10~K and \LaAlSi (b) at 10 and 300~K with $E\parallel a$ compared with calculated \OC spectra using \DFT and \DFTDMFT with CTQMC and HIA for \CeAlSi and \DFT for \LaAlSi.
In \LaAlSi at 10~K, the experimental \OC spectrum has a large peak appearing at $\sim0.4~{\rm eV}$, and two shoulder structures at $\sim0.15~{\rm eV}$ and $\sim0.65~{\rm eV}$.
At 300~K, these shoulder structures smear out, leaving only the 0.4-eV peak.
These three features observed at 10~K are also roughly reproduced by the \DFT calculations, namely the peak at 0.35~eV and the shoulder structures at 0.1~eV and 0.55~eV.
Consequently, the electronic structure of \LaAlSi can be reproduced by the itinerant calculation of \DFT.

On the other hand, for \CeAlSi shown in Fig.~\ref{fig:OC}(a), the experimental \OC spectrum revealed a shoulder structure at 0.25~eV and peak structures at 0.5 and 0.8~eV.
In contrast, the \OC spectrum obtained from \DFT calculations shows large peaks at 0.1 and 0.5~eV, which clearly differ from the experimental results.
The origin of these peaks at 0.5 eV lies in optical transitions from the valence band to itinerant unoccupied $4f$ levels assumed in \DFT.
Since these two peaks are not observed experimentally, this indicates that \CeAlSi exhibits weak itinerancy, that is, it is highly localized.

We therefore introduced \DFTDMFT calculations.
Both \OC spectra obtained from two kinds of \DFTDMFT calculations, with HIA and CTQMC, showed a shoulder structure at 0.25~eV and peak structures at 0.45~eV and 0.8~eV.
These structures correspond well to the three characteristic features observed experimentally, and the shape and peak positions in the \OC spectrum were well reproduced.
Therefore, in the following sections, we investigate how the Weyl electronic states are affected by the FM transition by comparing the experimental results with the \DFTDMFT calculations.
It should be noted that the \OC spectrum from \DFTDMFT with CTQMC is very similar to that with HIA, suggesting that the Ce~$4f$ electronic structure of \CeAlSi is almost localized.


\begin{figure}
\begin{center}
\includegraphics[width=0.48\textwidth]{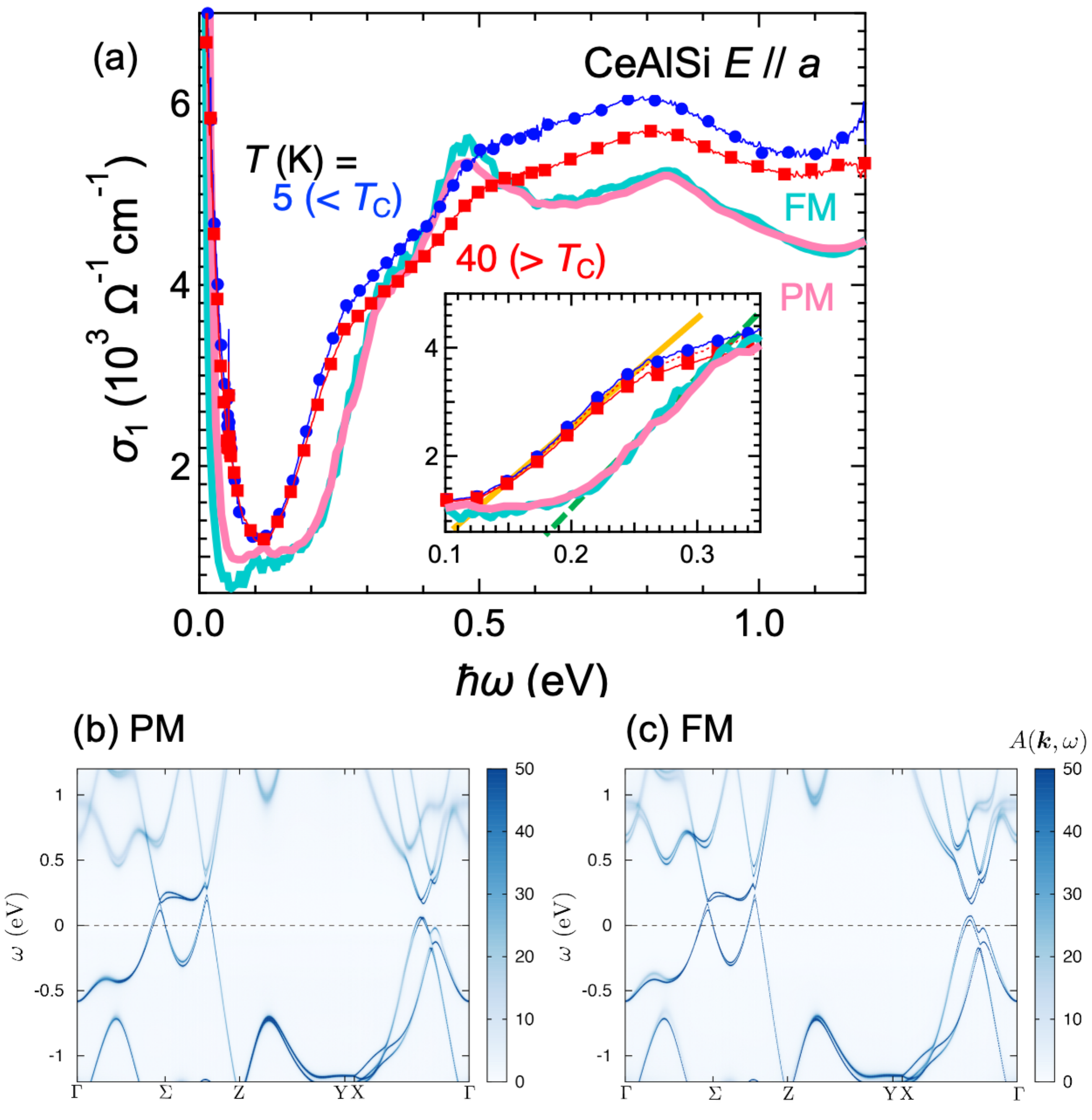}
\end{center}
\caption{
(a) Experimental \OC spectra of \CeAlSi with $E\parallel a$ in the PM phase (solid squares) at $T~=~40~{\rm K}$ and FM phase (solid circles) at $T~=~5~{\rm K}$ compared to the \OC spectra calculated with \DFTDMFT with CTQMC in both phases.
(Inset) Extended figure near the onset of the interband transition in the \hw region below 0.35~eV.
The solid and dashed lines show the onset slopes of the experimental and theoretical spectra, respectively, as a guide to the eye.
(b, c) Band structures in the PM phase (b) and the FM phase (c) of \CeAlSi corresponding to the calculated \OC spectra in (a).
}
\label{fig:TdepOC}
\end{figure}
Next, we discuss the spectral change across the FM ordering temperature.
Figure~\ref{fig:TdepOC}(a) shows the experimental \OC spectra in the PM phase ($T = 40~{\rm K}$) and the FM phase ($T = 5~{\rm K}$), and the corresponding theoretical spectra calculated by \DFTDMFT with CTQMC.
In the FM state, the Weyl band is expected to split due to a Zeeman splitting caused by the internal magnetic field, as described in PrAlSi~\cite{Wang2025-jm,Gao2025-pq}.
However, in \CeAlSi, although a slight increase in the intensity of the interband transition was observed below \TC, no spectral changes corresponding to a Zeeman splitting were observed.
Furthermore, the $\omega$-linear component expected from the Weyl band~\cite{Ashby2014-uf} was observed in the range of $0.13-0.23~{\rm eV}$ (see the inset of Fig.~\ref{fig:TdepOC}), and the $v_{\rm F}$ derived from this is evaluated as $(1.0 \pm 0.15) \times 10^5$~m/s; however, it shows almost no change across \TC.
This result is also reproduced in the \OC spectrum obtained from the \DFTDMFT calculations.
In other words, while the intensity of the \OC above 0.2~eV increases slightly in the FM phase, the $\omega$-linear component expected from the Weyl band remains virtually unchanged, with $v_{\rm F} = 0.93 \times 10^5$~m/s in the PM phase and $v_{\rm F} = 0.82 \times 10^5$~m/s in the FM phase.
These results suggest that the Weyl electronic states remain unchanged across \TC, as shown by both experimental data and \DFTDMFT calculations.
From the \DFTDMFT results for the PM and FM phases shown in Fig.~\ref{fig:TdepOC}(b, c), it can be seen that bands distant from \EF, such as those at $\sim0.9~{\rm eV}$ at the $\Gamma$ point and at $\sim-0.9$ and $-0.4~{\rm eV}$ on the $\Gamma-\Sigma$ line and so on, exhibit greater splitting in the FM phase, whereas the bands close to \EF are almost steady.
Therefore, the Weyl bands of \CeAlSi remain virtually unchanged at the FM transition.
However, the starting energy of the slope in the calculation shifts by about 0.1~eV relative to the experimental value, which originates from the energy position of the Fermi level ($E_{\rm F}$)~\cite{Tabert2016-ck}, i.e., the Weyl point of the experimentally used sample is closer to \EF than that of the calculated band structure.

It should be noted that the $\omega$-linear behavior of \LaAlSi is shown in the inset of Fig.~\ref{fig:OC}(b).
According to this, \DFT predicts $\omega$-linearity in the range of $0.04-0.1~{\rm eV}$, and experimentally it appears in almost the same range of $0.04-0.14~{\rm eV}$ at $10~{\rm K}$.
However, the slope differs by about a factor of two.
This implies that the actual Fermi velocity, $v_{\rm F} = dE/(\hbar dk)\vert_{k=k_{\rm F}}$, which is the dispersion of the Weyl bands at \EF, is about half that predicted by the calculation.
Furthermore, at 300~K, the 0.14-eV shoulder structure observed at 10~K smeared out, and a $\omega$-linear behavior extending to 0.24~eV appeared.
Since this structure can be regarded as the averaged spectral structure from \DFT calculations, it is thought to be an interband transition within the Weyl band that appears due to temperature-induced broadening.
This $\omega$-linear component has been observed in previously reported PrAlSi and NdAlSi~\cite{Kunze2024-qf} and LaAlGe and CeAlGe~\cite{Corasaniti2021-ta}; the actual Weyl cone component should be clarified, possibly at low temperatures.


\begin{figure}
\begin{center}
\includegraphics[width=0.48\textwidth]{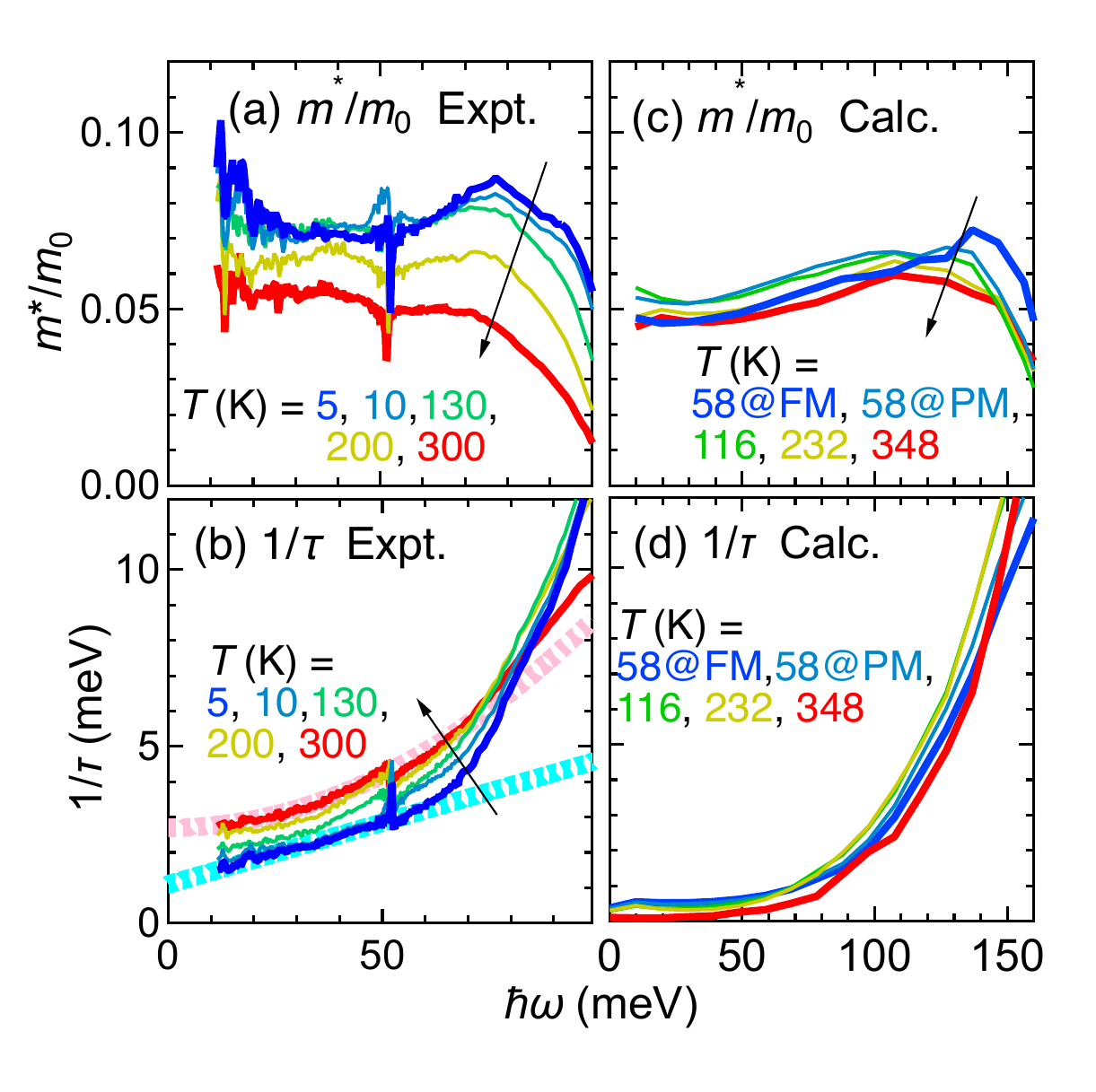}
\end{center}
\caption{
The experimentally obtained temperature-dependent dynamical effective mass of carriers relative to the electron rest mass ($m^*/m_0$, a) and scattering rate ($1/\tau$, b) of \CeAlSi compared with the theoretical $m^*/m_0$ (c) and $1/\tau$ (d) calculated with \DFTDMFT.
In (c) and (d), 58@FM (58@PM) indicates the calculation with the FM (PM) magnetic structure at $T = 58~{\rm K}$. 
The dashed lines in (b) indicate the $\omega$-linear and $\omega$-square behavior for $T = 5$ and $300~{\rm K}$, respectively, as a guide to the eye.
}
\label{fig:EDA}
\end{figure}
Finally, we analyzed the properties of the conduction band using the extended Drude model~\cite{Dressel2002-ae}.
To evaluate the effective mass, we used the carrier density value ($1 \times 10^{20}~{\rm cm}^{-3}$) obtained from the Hall effect~\cite{Piva2023-ij}.
As shown in Fig.~\ref{fig:EDA}, the ratio of the effective mass of the carriers to the rest mass of an electron ($m^*/m_0$) at the lowest accessible photon energy of 10~meV increased from $\sim 0.05$ at 300~K to $\sim 0.08$ at temperatures below 10~K.
The value at low temperatures is consistent with that evaluated with the Shubnikov-de Haas effect~\cite{Meena2025-pj}.
This very-low effective mass is in contrast to typical heavy-electron systems, where a mass increase of more than one order of magnitude has been observed~\cite{Kimura2006-pd}.
This result suggests that the Ce~$4f$ states are almost localized in this material.
The spectral shapes of $m^*/m_0$ at low and high temperatures are very similar to those calculated with \DFTDMFT shown in Fig.~\ref{fig:EDA}(c);
At temperatures below 10~K, there is a hump at $\sim75~{\rm meV}$, which corresponds to the peak at around 140~meV in the calculated spectra.

The experimentally obtained scattering probability spectra ($1/\tau$) in Fig.~\ref{fig:EDA}(b) are also very similar to those of the calculation (Fig.~\ref{fig:EDA}(d)) except for the offset intensity at \hw$=0~{\rm meV}$; the upturn appears higher than 70~meV.
In the experimental spectra, $1/\tau$ exhibits Fermi-liquid-like behavior proportional to $\omega^2$ at $T=300~{\rm K}$; however, at 5~K, despite being in the FM state, it follows an $\omega$-linear behavior for \hw~$\leq 50~{\rm meV}$.
This possibly suggests that magnetic fluctuations might be present in the FM phase~\cite{Kimura2025-tx}.
The magnetic moment of this material is $1.7~\mu_{\rm B}$ ~\cite{Meena2025-pj}, which slightly deviates from the value expected from the fully localized magnetism in Ce~$4f^1$.
This is consistent with the fact that the \OC spectra of \CeAlSi cannot be explained by \DFT, but are well explained by \DFTDMFT.

%
To summarize, we investigated changes in the bulk electronic states during the magnetic transition by measuring the temperature dependence of the \OC spectrum of \CeAlSi and \LaAlSi and comparing them with band calculations.
The \OC spectrum of \LaAlSi could be reproduced with the \DFT calculations, suggesting the itinerant Weyl electronic structure.
The \OC of \CeAlSi in the PM phase, on the other hand, could not be explained by itinerant \DFT calculations, but was well reproduced by nearly localized \DFTDMFT calculations.
Across the FM ordering temperature, the overall spectral shape of the \OC spectrum did not change, but the intensity increased slightly from the PM to FM phase.
This slight spectral change due to the phase transition was also reproduced by \DFTDMFT.
The effective mass of the carriers of the Drude component was evaluated as $\sim0.05~m_0$ at room temperature and $\sim0.08~m_0$ at low temperatures, and there is a hump structure at \hw$\sim75~{\rm meV}$ at low temperatures, which could also be qualitatively reproduced by \DFTDMFT.
All of these results suggest that the Ce~$4f$ states in \CeAlSi are nearly localized and have almost no effect on the Weyl electronic structure.

%
We thank UVSOR Synchrotron staff members for their support during synchrotron radiation experiments.
Part of this work was performed under the Use-of-UVSOR Synchrotron Facility Program (Proposals No.~24IMS6018, No.~25IMS6014) of the Institute for Molecular Science, National Institutes of Natural Sciences.
Part of the computations was performed using the facilities of the Supercomputer Center, the Institute for Solid State Physics, the University of Tokyo (Grants No.~2025-Ca-0086 and No.~2026-Ca-0142).
This work was partly supported by JSPS KAKENHI (Grant No.~23H00090, No.~23H04869, No.~24K21197, No.~26K00659).

%
%

\bibliographystyle{apsrev4-1}
\bibliography{CeAlSi_Notes.bib,dmft.bib}

\end{document}



\title{
Supplementary material for ``Optical investigation on the electronic structure of a ferromagnetic Weyl semimetal CeAlSi''
}
    
\author{Shin-ichi~Kimura}
\email{sk@kimura-lab.com}
\affiliation{Graduate School of Frontier Biosciences, The University of Osaka, Suita, Osaka 565-0871, Japan}
\affiliation{Department of Physics, Graduate School of Science, The University of Osaka, Toyonaka, Osaka 560-0043, Japan}
\affiliation{Institute for Molecular Science, 
Okazaki, Aichi 444-8585, Japan}
\author{Yue~Pan}
\affiliation{Department of Physics, Graduate School of Science, The University of Osaka, Toyonaka, Osaka 560-0043, Japan}
\author{Hiroshi~Watanabe}
 \altaffiliation[Present affiliation: ]{Institute for Chemical Research, Kyoto University, Uji, Kyoto 611-0011, Japan}
\affiliation{Graduate School of Frontier Biosciences, The University of Osaka, Suita, Osaka 565-0871, Japan}
\affiliation{Department of Physics, Graduate School of Science, The University of Osaka, Toyonaka, Osaka 560-0043, Japan}
\author{Akimitsu~Kirikoshi}
\affiliation{Research Institute for Interdisciplinary Science, Okayama University, Okayama 700-8530, Japan}
\author{Junya~Otsuki}
\affiliation{Research Institute for Interdisciplinary Science, Okayama University, Okayama 700-8530, Japan}
\author{Hiroshi~Tanida}
\affiliation{Liberal Arts and Sciences, Toyama Prefectural University, Imizu, Toyama 939-0398, Japan}
\date{\today}
\maketitle
\setcounter{equation}{0}
\setcounter{figure}{0}
\setcounter{table}{0}
\setcounter{section}{0}
\setcounter{page}{1}
\makeatletter
\renewcommand{\thesection}{S\arabic{section}}
\renewcommand{\theequation}{S\arabic{equation}}
\renewcommand{\thefigure}{S\arabic{figure}}
\renewcommand{\bibnumfmt}[1]{[S#1]}
\renewcommand{\citenumfont}[1]{S#1}

\section{Derivation of optical conductivity spectra from near-normal-incident reflectivity spectra}

%
Near-normal-incident reflectivity [$R(\omega)$] spectra were acquired over a wide \hw range of 10~meV--30~eV to ensure accurate Kramers-Kronig analysis (KKA)~\cite{Kimura2013-rg}.
Spectroscopic measurements at \hw = 10~meV--1.2~eV at temperatures of 5--300~K have been performed using \R measurement setups with Michelson-type FTIR spectrometers (FT/IR-6100, JASCO Co.), combined with several light sources (Ceramic and halogen lamps), beam splitters (Quartz, Ge/KBr, and multilayered mylar film), and detectors (MCT and Si bolometer).
The absolute values of \R spectra were determined with the {\it in-situ} gold evaporation method.
In the \hw range of 1.5--30~eV, the \R spectrum was acquired only at 300~K by using the synchrotron radiation setup at the beamline 3B with a 2.5-m normal-incident VUV monochromator (HOTALU)~\cite{Fukui2014-wz} of UVSOR-III Synchrotron~\cite{Ota2022-ak}, and was connected to the temperature-dependent \R spectra for \hw $\leq 1.2$~eV for conducting accurate KKA.
Obtained temperature-dependent \R spectra of \CeAlSi are shown in Fig.~1(a1), and the wide-energy-range \R spectrum at 300~K is shown in Fig.~1(a2).
In order to obtain \OC via KKA of \R, the spectra were extrapolated below 10~meV with a Hagen-Rubens function [$R(\omega)=1-\{2\omega/(\pi\sigma_{DC})\}^{1/2}$] due to the metallic \R spectra, and above 30~eV with a free-electron approximation $R(\omega) \propto \omega^{-4}$ ~\cite{Dressel2002-ae}.
Here, the direct current conductivity ($\sigma_{DC}$) values were adopted from the experimental values \cite{Piva2025-oj}.
The extrapolations were confirmed not to severely affect the \OC spectra at \hw$\sim0.01-1$~eV, which is the main part of this paper.
The derived \OC spectra are shown in Figs.~1(b1) and 1(b2) for the temperature dependence of the wide energy range and the TO-phonon peak, respectively, and in Fig.~1(b3) for the higher energy range only at 300~K.

\section{Reflectivity spectra with different surface treatments}
\begin{figure*}[b]
\includegraphics[width=0.8\textwidth]{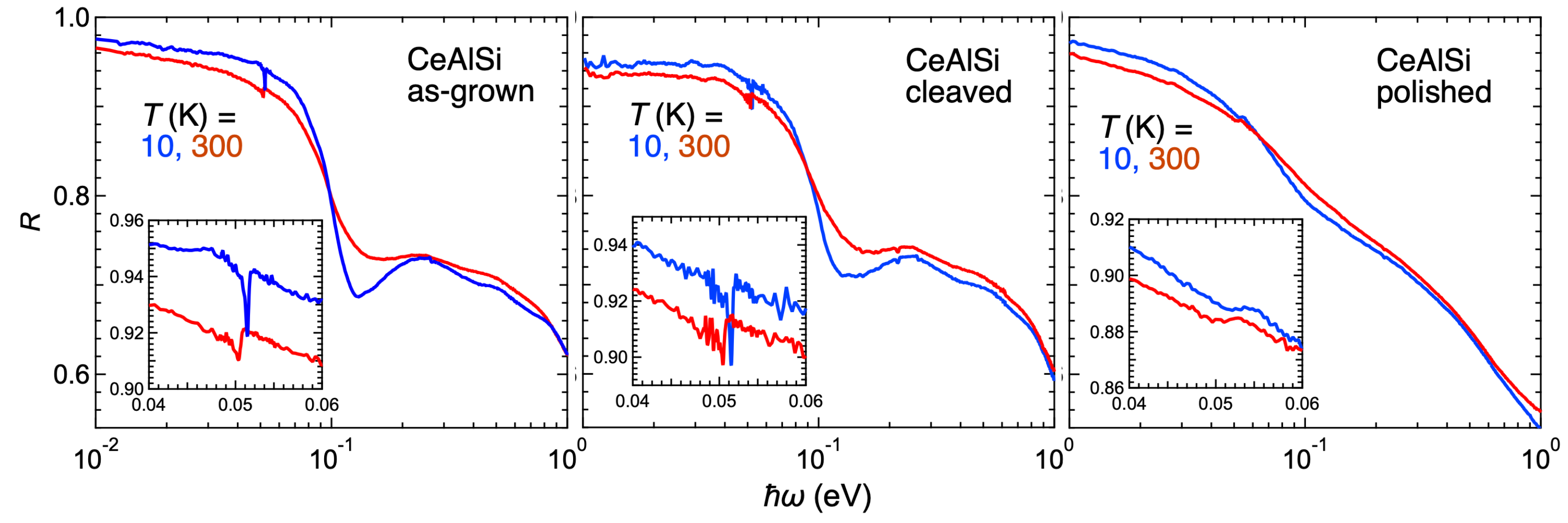}
\caption{Reflectivity spectra of \CeAlSi at temperatures of 10 and 300~K with different surface treatments: (a) as grown surface with the size of $\sim4~{\rm mm}^{2}$, (b) cleaved srface with the size of $\sim1~{\rm mm}^{2}$, (c) polished surface with the size of $\sim4~{\rm mm}^{2}$ with 3M$^{\rm TM}$ Lapping Film Sheets (0.3-Micron Grade).
The insets indicate the TO-phonon absorption peaks.
}
\label{fig:supfig1}
\end{figure*}

To investigate the effect of surface treatment on the \R spectrum of \CeAlSi, we measured the \R spectra with $E \parallel a$ of an as-grown surface, a cleaved surface, and a polished surface, as shown in Fig.~\ref{fig:supfig1}.
Samples from the same batch were used for these measurements.
Although the as-grown surface and the cleaved surface exhibit nearly identical spectral shapes, the polished surface shows a flattened spectral shape.
Furthermore, the optical phonon structure (shown in the inset) of the polished surface is clearly degraded and smaller compared to that of the as-grown and cleaved surfaces.
This suggests that the material within the depth range observed in optical measurements (Inverse absorption coefficient $1/\alpha \geq 100~{\rm nm}$) differs from that of the bulk \CeAlSi due to polishing.
On the other hand, the fact that the as-grown and cleaved surfaces exhibit nearly identical shapes suggests that no Al flux or other residues remain on the surface of the as-grown sample.
Although the cleaved surface is likely to correspond to the bulk properties, the flat area on the cleaved surface where optical measurements are possible is only about 1 mm$^2$; since a sufficient signal-to-noise (S/N) ratio cannot be obtained, this study measured the optical reflection spectra on the as-grown surface, which is approximately 4 mm$^2$ and allows a sufficient S/N ratio, and used these results for the discussion in this study.

\section{Band structure and Weyl points in the PM and FM phases calculated with \DFTDMFT}

\begin{figure*}
\includegraphics[width=0.9\linewidth]{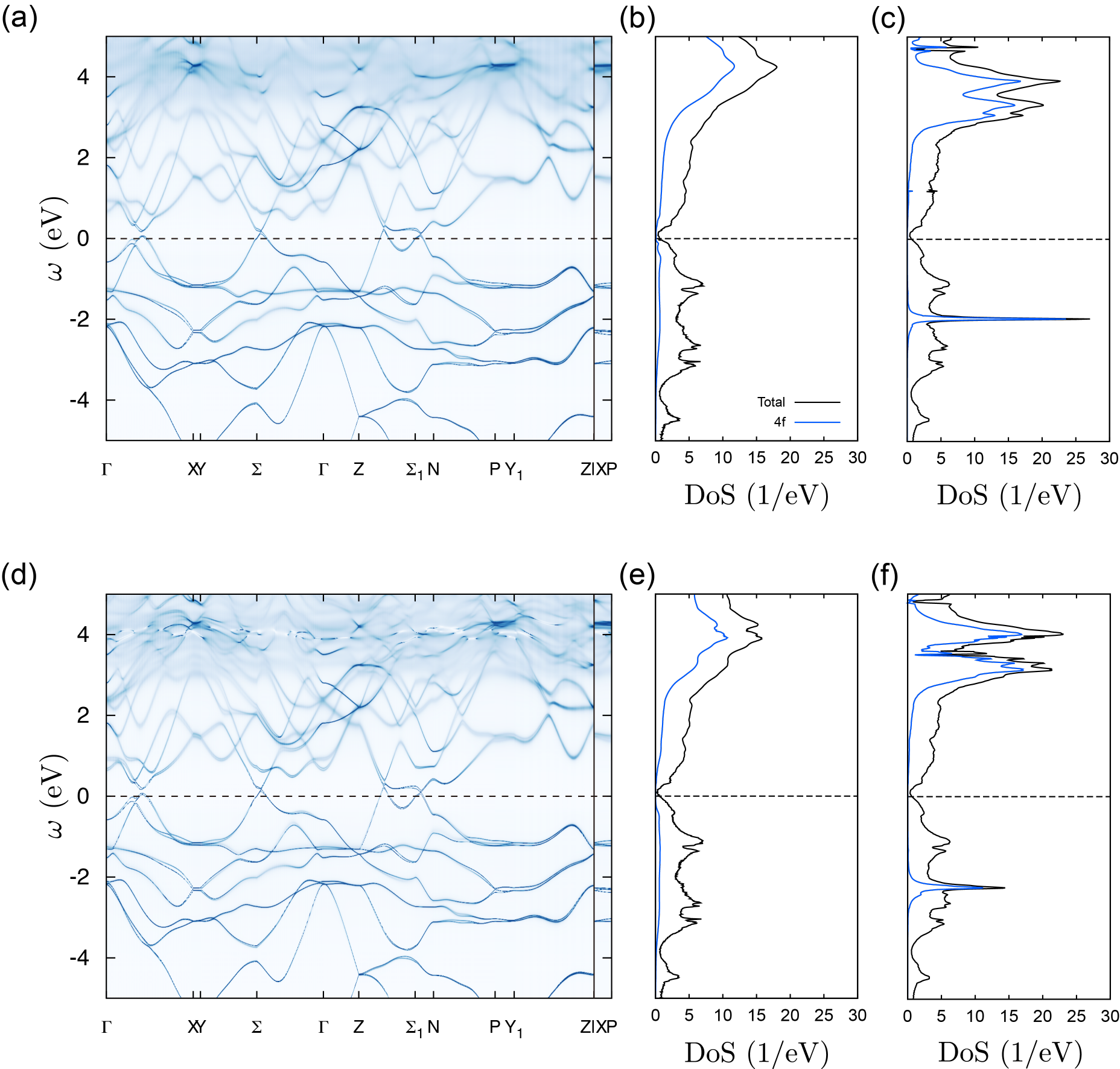}
\caption{
(a) Single-particle excitation spectrum $A(\bm{k},\omega)$ and (b) corresponding density of states $A(\omega)$ calculated with \DFTDMFT in the PM phase. (c) $A(\omega)$ calculated within the Hubbard-I approximation in the PM phase. (d)-(f) Corresponding results for the FM phase, with the panels arranged as in (a)-(c).
}
\label{fig:supfig2}
\end{figure*}
\begin{figure*}
\includegraphics[width=0.25\textwidth]{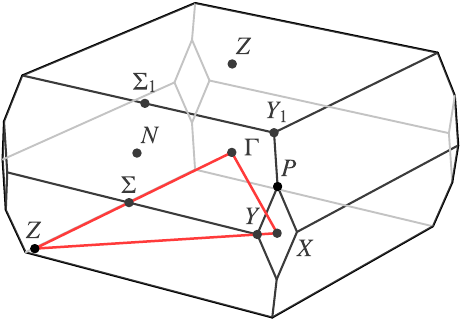}
\caption{
The Brillouin zone and high-symmetry points of CeAlSi. The red line indicates the path adopted in Fig.~3(b, c).
}
\label{fig:bz}
\end{figure*}

Figure~\ref{fig:supfig2}(a) shows the single-particle excitation spectrum $A(\bm{k},\omega)$ in the PM phase calculated in the \DFTDMFT method. 
The high-symmetry points are defined in Fig.~\ref{fig:bz}.
A characteristic linear dispersion is seen near \EF as discussed in Fig.~3(b). 
Figure~\ref{fig:supfig2}(b) shows the corresponding density of states $A(\omega)$, and for comparison, the result obtained within the Hubbard-I approximation is shown in Fig.~\ref{fig:supfig2}(c).
The $4f^0$ peak is located around $-2.0$~eV, and the lowest $4f^2$ multiplet appears around $3.1$~eV. 
These peaks are broadened in the \DFTDMFT results due to hybridization with conduction electrons.
The results for the FM phase are shown in Figs.~\ref{fig:supfig2}(d)--(f).
The $4f$ peaks are split by the FM molecular field, resulting in a reduction of the intensity of the $4f^0$ peak around $-2.0$~eV.
This splitting affects the conduction bands near \EF, as demonstrated in Fig.~3(c).

\begin{figure*}
\includegraphics[width=0.8\linewidth]{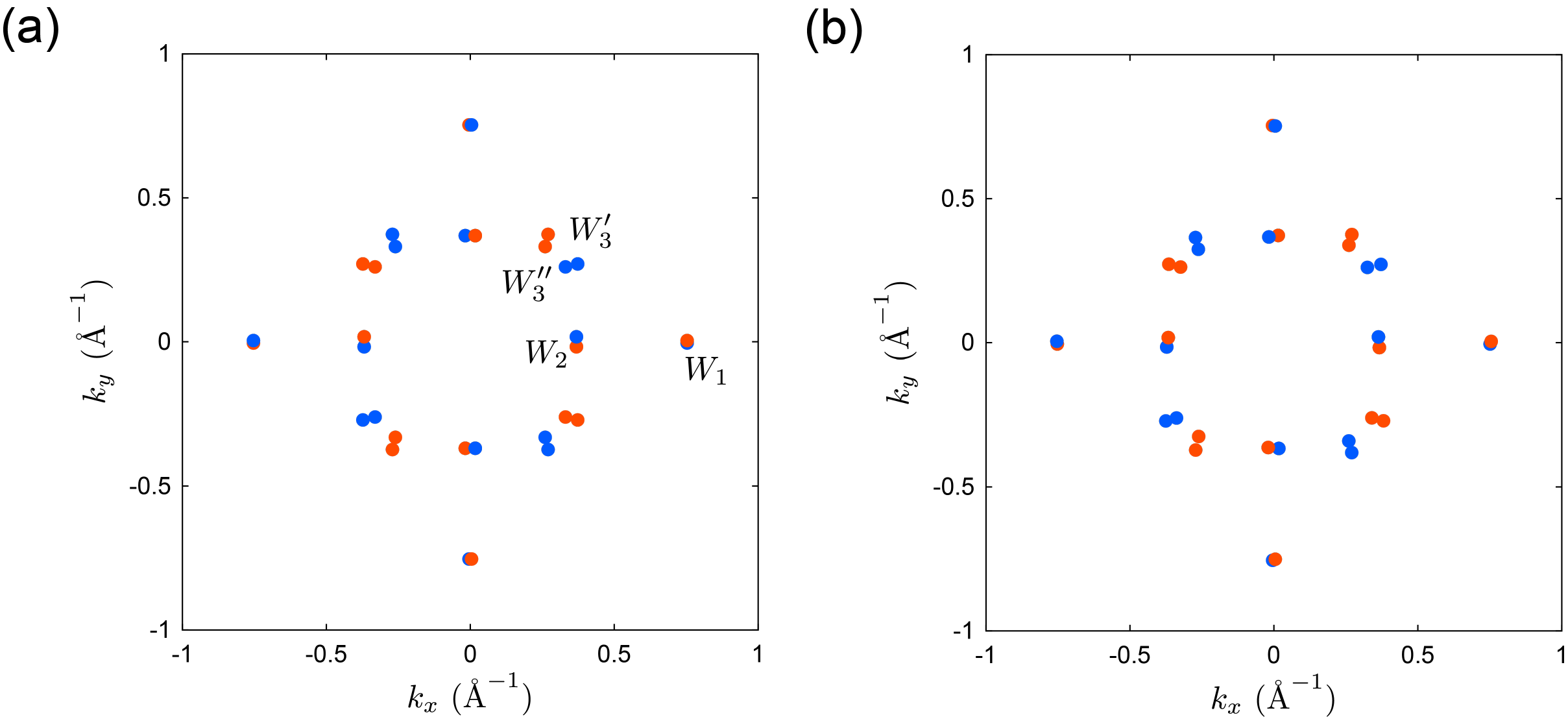}
\caption{
Distribution of the Weyl points projected onto the $[001]$ surface in the Brillouin zone for (a) the PM phase and (b) the FM phase. The red and blue points indicate $\bm{k}^+$ and $\bm{k}^-$ with chirality $+1$ and $-1$, respectively.
}
\label{fig:Weyl_projection}
\end{figure*}

Table~\ref{tab:Weyl_nodes_PM} summarizes the momenta and energies of the Weyl points in the PM phase.
The Weyl-point search was performed using WannierTools~\cite{WU2018} after constructing an effective Hamiltonian that includes local correlations within the Hubbard-I approximation~\cite{kirikoshi2026nonlocalkondoexchangedrivenintrinsicanomalous}.
The search focused on intersections between the band crossing \EF and the unoccupied band just above it, and 
identified 40 Weyl points, which are consistent with the previous work~\cite{Yang2021-ht,Sakhya2023-en,Morita2025-ek}.
Weyl points connected by the mirror symmetry $m_{[110]}$ have opposite chiralities.
$W_1$ and $W_2$ comprise four and eight symmetry-equivalent pairs, respectively, related by fourfold rotational symmetry or the product of the twofold rotation $C_{2z}$ and time reversal.
$W_{3}^{\prime}$ and $W_{3}^{\prime\prime}$ are Weyl nodes due to the splitting by the spin--orbit coupling~\cite{Sakhya2023-en}.
The Weyl points in the FM phase are listed in Table~\ref{tab:Weyl_nodes_FM}.
In contrast to the PM case, the fourfold symmetry of the Weyl points is lifted in both energy and momentum.
The distributions of the Weyl points projected onto the $[001]$ surface are shown in Figs.~\ref{fig:Weyl_projection}(a) and \ref{fig:Weyl_projection}(b) for PM and FM states, respectively.
The small shifts in the momenta can be confirmed, for example, for $W_2$ around $(k_x, k_y) \approx (0.37, \pm 0.02)~\mathrm{\AA}^{-1}$.


\begin{table*}[hb]
    \centering
    \begin{tabular}{c|c|c}
        \hline
         Label & $\bm{k}^{+},\bm{k}^{-}~(\mathrm{\AA}^{-1})$ & $E-E_{F}~(\mathrm{meV})$ \\
         \hline
         $W_{1}$ & $(+0.753,+0.004,0),(-0.004,-0.753,0)$ & $174.0$ \\
         & $(-0.004,+0.753,0),(-0.753,+0.004,0)$ & \\
         & $(-0.753,-0.004,0),(+0.004.+0.753,0)$ & \\
         & $(+0.004,-0.753,0),(+0.753,-0.004,0)$ & \\
         \hline
         $W_{2}$ & $(+0.017,+0.369,\pm0.288),(-0.369,-0.017,\pm0.288)$ & $138.5$ \\
         & $(+0.369,-0.017,\pm0.288),(+0.017,-0.369,\pm0.288)$ & \\
         & $(-0.369,+0.017,\pm0.288),(-0.017,+0.369,\pm0.288)$ & \\
         & $(-0.017,-0.369,\pm0.288),(+0.369,+0.017,\pm0.288)$ & \\
         \hline
         $W_{3}^{\prime}$ & $(-0.373,+0.270,0),(-0.270,+0.373,0)$ & $144.7$ \\
         & $(-0.270,-0.373,0),(+0.373,+0.270,0)$ & \\
         & $(+0.270,+0.373,0),(-0.373,-0.270,0)$ & \\
         & $(+0.373,-0.270,0),(+0.270,-0.373,0)$ & \\
         \hline
         $W_{3}^{\prime\prime}$ & $(+0.331,-0.260,0),(+0.260,-0.331,0)$ & $131.5$ \\
         & $(+0.260,+0.331,0),(-0.331,-0.260,0)$ & \\
         & $(-0.260,-0.331,0),(+0.331,+0.260,0)$ & \\
         & $(-0.331,+0.260,0),(-0.260,+0.331,0)$ & \\
         \hline
    \end{tabular}
    \caption{Momenta and energies of the Weyl points in the PM phase.
    Each row contains a pair of Weyl points connected by the mirror symmetry $m_{[110]}$.
    $\bm{k}^{\pm}$ represents the momentum with chirality $\chi=\pm1$. 
    The convention of the labels follows Ref.~\cite{Yang2021-ht}.}
    \label{tab:Weyl_nodes_PM}
\end{table*}

\begin{table*}[hb]
    \centering
    \begin{tabular}{c|c|c}
        \hline
         Label & $\bm{k}^{+},\bm{k}^{-}~(\mathrm{\AA}^{-1})$ & $E-E_{F}~(\mathrm{meV})$ \\
         \hline
         $W_{1}^{1}$ & $(+0.755,+0.004,0),(-0.004,-0.755,0)$ & $172.4$ \\
         $W_{1}^{2}$ & $(-0.005,+0.754,0),(-0.754,+0.005,0)$ & $174.4$ \\
         $W_{1}^{3}$ & $(-0.752,-0.005,0),(+0.005,+0.752,0)$ & $175.7$ \\
         $W_{1}^{4}$ & $(+0.005,-0.751,0),(+0.751,-0.005,0)$ & $180.7$ \\
         \hline
         $W_{2}^{1,5}$ & $(+0.015,+0.372,\pm0.287),(-0.372,-0.015,\pm0.287)$ & $127.3$ \\
         $W_{2}^{2,6}$ & $(+0.366,-0.017,\pm0.291),(+0.017,-0.366,\pm0.291)$ & $136.0$ \\
         $W_{2}^{3,7}$ & $(-0.367,+0.017,\pm0.291),(-0.017,+0.367,\pm0.291)$ & $136.7$ \\
         $W_{2}^{4,8}$ & $(-0.020,-0.363,\pm0.293),(+0.363,+0.020,\pm0.293)$ & $144.0$ \\
         \hline
         $W_{3}^{\prime1}$ & $(-0.365,+0.272,0),(-0.272,+0.365,0)$ & $138.5$ \\
         $W_{3}^{\prime2}$ & $(-0.272,-0.372,0),(+0.372,+0.272,0)$ & $141.6$ \\
         $W_{3}^{\prime3}$ & $(+0.271,+0.375,0),(-0.375,-0.271,0)$ & $147.0$ \\
         $W_{3}^{\prime4}$ & $(+0.381,-0.270,0),(+0.270,-0.381,0)$ & $151.5$ \\
         \hline
         $W_{3}^{\prime\prime1}$ & $(+0.340,-0.260,0),(+0.260,-0.340,0)$ & $117.7$ \\
         $W_{3}^{\prime\prime2}$ & $(+0.261,+0.338,0),(-0.338,-0.261,0)$ & $120.7$ \\
         $W_{3}^{\prime\prime3}$ & $(-0.261,-0.325,0),(+0.325,+0.261,0)$ & $137.2$ \\
         $W_{3}^{\prime\prime4}$ & $(-0.324,+0.262,0),(-0.262,+0.324,0)$ & $138.9$ \\
         \hline
    \end{tabular}
    \caption{Momenta and energies of the Weyl points in the FM phase. See also the caption of Table~\ref{tab:Weyl_nodes_PM}.}
    \label{tab:Weyl_nodes_FM}
\end{table*}

%
%
\bibliographystyle{apsrev4-1}
\bibliography{CeAlSi_Notes,dmft}